\documentclass[english]{article}
\usepackage[T1]{fontenc}
\usepackage[latin9]{inputenc}
\usepackage{geometry}
\usepackage{color}
\usepackage{float}
\usepackage{graphicx}
\usepackage{amsmath}
\usepackage{amssymb}
\usepackage{babel}


\usepackage{babel}
\begin{document}

\title{Comment on Quantum teleportation via GHZ-\emph{like} state}

\author{Anindita Banerjee, Kamal Patel and Anirban Pathak}

\maketitle
\begin{center}
Jaypee Institute of Information Technology University, Noida, India
\par\end{center}

%%%%%%%%%%%%%%%%%%%%%%%%%%%%%%%%%%%%%%%%%%%%%%%%%%%%%%%%%%%%%%%%%%%%%%%%%%%%%%%

\begin{abstract}
Recently Yang \emph{et al.} {[}Int. J. Theo. Phys. \textbf{48} (2009) 516{]} have shown that an unknown qubit
can be teleported by using a particular GHZ-\emph{like} state as quantum channel. However, there are several
errors in the calculation which lead to incorrect conclusions. The errors have been indicated and corrected. It
is also noted that their scheme and the independently proposed teleportation scheme of Zhang \emph{et al.}
{[}Int. J. Theo. Phys. \textbf{48} (2009) 3331{]} uses quantum channel from the same family and any state of
that family may be used for teleportation.
\end{abstract}

%%%%%%%%%%%%%%%%%%%%%%%%%%%%%%%%%%%%%%%%%%%%%%%%%%%%%%%%%%%%%%%%%%%%%%%%%%%%%%%%%%%%%%%5

In a recent paper Yang \emph{et al.} \cite{yang-c} have shown that an unknown qubit can be teleported by using
GHZ-\emph{like} state as quantum channel. In \cite{yang-c} Yang \emph{et al.} have made some mistakes in
calculation. For example, in section 2.1 of \cite{yang-c} they start preparing their channel with a quantum
state $|\phi\rangle_{1}=|0\rangle_{1}+|1\rangle_{1}$ which is not normalized. Ideally it should read as
$|\phi\rangle_{1}=\frac{1}{\sqrt{2}}\left(|0\rangle_{1}+|1\rangle_{1}\right)$. This is important because without
this normalization constant the third line of equation (1) of \cite{yang-c} will not follow from the second line
of the same equation. As $|\phi\rangle_{1}$ can be produced by operating Hadamard transformation $H$ on the
state $|0\rangle,$ the quantum circuit that prepares the GHZ-\emph{like} quantum channel used by Yang \emph{et
al.} is essentially two EPR circuits attached in sequence in the cascaded manner (see Fig. 1).
\begin{figure}
\centering \includegraphics[scale=0.7]{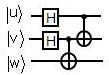} \caption{A circuit to create GHZ-\emph{like} states
$\left|u\right\rangle
$$\left|v\right\rangle \left|w\right\rangle $ are the input bits to the circuit and the output is a
GHZ-\emph{like} state} \label{com-fig1}
\end{figure}

After preparing the GHZ\emph{-like} quantum channel they use that channel to teleport an unknown qubit
$\alpha|0\rangle+\beta|1\rangle$. To do so Alice keeps the first two qubits of the channel with herself (i.e.
particle 1 and 2) and particle 3 and 4 are sent to Bob and Charlie respectively. Now the product state of the
unknown qubit and the channel is

\begin{equation}
\begin{array}{lcl}
|\psi\rangle_{1234} & = & (\alpha|0\rangle+\beta|1\rangle)_{1}\otimes\frac{1}{2}(|001\rangle+|010\rangle+|100\rangle+|111\rangle)_{234}\\
 & = & \frac{1}{2}(\alpha|000\rangle+\beta|100\rangle+\alpha|011\rangle+\beta|111\rangle)_{123}|1\rangle_{4}\\
 & + & \frac{1}{2}(\alpha|001\rangle+\beta|101\rangle+\alpha|010\rangle+\beta|110\rangle)_{123}|0\rangle_{4}.\\
 & = & |A\rangle_{123}|1\rangle_{4}+|B\rangle_{123}|0\rangle_{4}\end{array}\label{eq:1}\end{equation}

This equation coincides with equation (5) of \cite{yang-c}. After this step they have tried to expand
$|A\rangle_{123}$ and $|B\rangle_{123}$ as sum of product of Bell states (involving particle 1 and 2) and one
qubit state so that if Charlie measures his qubit in computational basis and Alice does a Bell measurement on
her qubits, then the necessary information about the unknown qubit is transferred to Bob. The outcome of the
measurements are sent to Bob via classical channel and after receiving those information Bob chooses a suitable
unitary operation to recreate the unknown state. Yang \emph{et al.} did simple mistakes in the expansion step.
The last two lines of both equations (6) and (7) of \cite{yang-c} are incorrect and consequently the unitary
operations reported in Table 1 of \cite{yang-c} which are supposed to be performed by Bob are also incorrect.
Correct expansion of product state would be \begin{equation}
\begin{array}{lcl}
|\psi\rangle_{1234} & = & \frac{1}{2\sqrt{2}}\left[|\psi^{+}\rangle_{12}(\alpha|0\rangle+\beta|1\rangle)_{3}+|\psi^{-}\rangle_{12}(\alpha|0\rangle-\beta|1\rangle)_{3}\right.\\
 & + & \left.|\phi^{+}\rangle_{12}(\alpha|1\rangle+\beta|0\rangle)_{3}+|\phi^{-}\rangle_{12}(\alpha|1\rangle-\beta|0\rangle)_{3}\right]|1\rangle_{4}\\
 & + & \frac{1}{2\sqrt{2}}\left[|\psi^{+}\rangle_{12}(\alpha|1\rangle+\beta|0\rangle)_{3}+|\psi^{-}\rangle_{12}(\alpha|1\rangle-\beta|0\rangle)_{3}\right.\\
 & + & \left.|\phi^{+}\rangle_{12}(\alpha|0\rangle+\beta|1\rangle)_{3}+|\phi^{-}\rangle_{12}(\alpha|0\rangle-\beta|1\rangle)_{3}\right]|0\rangle_{4}\end{array}.\label{eq:2}\end{equation}
 Using (2) one can easily generate the Table 1 and this table considerably
differs from the Table 1 of \cite{yang-c}.

\begin{table}
\centering  \caption{Corrected table indicating the operations to be performed by Bob to construct unknown state
at his port.}\label{com-table1}
\begin{tabular}{lll}
\hline\noalign{\smallskip}
Outcomes of Alice & Outcomes of Charlie & Unitary operation of Bob\\
\noalign{\smallskip}\hline\noalign{\smallskip}
 $|\psi^{+}\rangle_{12}$ & $|1\rangle_{4}$ &$I=\left(\begin{array}{cc}1 & 0\\
0 & 1\end{array}\right)$\\
 $|\psi^{-}\rangle_{12}$ & $|1\rangle_{4}$ & $Z=\left(\begin{array}{cc}
1 & 0\\
0 & -1\end{array}\right)$\\
$|\phi^{+}\rangle_{12}$ & $|1\rangle_{4}$ & $X=\left(\begin{array}{cc}
0 & 1\\
1 & 0\end{array}\right)$\\
$|\phi^{-}\rangle_{12}$ & $|1\rangle_{4}$ & $ZX=\left(\begin{array}{cc}
0 & 1\\
-1 & 0\end{array}\right)$\\
 $|\psi^{+}\rangle_{12}$ & $|0\rangle_{4}$ & $X=\left(\begin{array}{cc}
0 & 1\\
1 & 0\end{array}\right)$\\
 $|\psi^{-}\rangle_{12}$ & $|0\rangle_{4}$ & $ZX=\left(\begin{array}{cc}
0 & 1\\
-1 & 0\end{array}\right)$\\
 $|\phi^{+}\rangle_{12}$ & $|0\rangle_{4}$ & $I=\left(\begin{array}{cc}
1 & 0\\
0 & 1\end{array}\right)$\\
 $|\phi^{-}\rangle_{12}$ & $|0\rangle_{4}$ & $Z=\left(\begin{array}{cc}
1 & 0\\
0 & -1\end{array}\right)$\\
\noalign{\smallskip}\hline
\end{tabular}
\end{table}
The erroneous calculation continues to the fidelity calculation. In this section they assume
\begin{enumerate}
\item Charlie has not measured the particle 4 but Alice has done the Bell measurement. \item After the Bell
measurement the state of particle 3 and 4 are in the state (see equation (9) of \cite{yang-c}).
\begin{equation}
|\phi\rangle_{34}=(\alpha|01\rangle+\beta|11\rangle)_{34}=(\alpha|0\rangle+\beta|1\rangle)_{3}|1\rangle_{4}.\label{eq:3}\end{equation}

\end{enumerate}
This is impossible and one can not bring particle 3 and 4 in a separable state without doing any measurement on
any one of them. Further, since in the equation (9) of \cite{yang-c}, the state of Bob (particle 3) is already
in the desired state ($\alpha|0\rangle+\beta|1\rangle)$. The Fidelity should be 1. They have really got it 1 as
in their calculation $F_{3}=(|\alpha|^{2}+|\beta|^{2})^{2}=1$ (see equation (12) of \cite{yang-c}). But for some
unclear reasons they could not recognize the fidelity as unity and they choose non physical condition like
$|\alpha|=|\beta|=\frac{1}{2}$ which does not satisfy the fundamental relation $|\alpha|^{2}+|\beta|^{2}=1$.
These non physical considerations lead to the conclusion that Fidelity varies between $\frac{1}{2}$ and 1 and
rest of the conclusions of Yang \emph{et al.} follows from this incorrect conclusion. Actually if we assume that
Charlie is not cooperating in this controlled teleportation scheme and he has not measured his bit and Alice has
done Bell measurement on her qubit then the corrected scenario will be as described in Table 2.

\begin{table}
\centering  \caption{Fidelity of Bob's state when Charlie is noncooperative.}\label{com-table2}
\begin{tabular}{lll}
\hline \noalign{\smallskip}
 Alice's Bell measurement & Joint state of Bob and Charlie & Fidelity
($F_{3})$\\
\noalign{\smallskip}\hline\noalign{\smallskip}
 $|\psi^{+}\rangle_{12}$ &
$\frac{1}{\sqrt{2}}\left(\beta|00\rangle+\alpha|01\rangle+\alpha|10\rangle+\beta|11\rangle\right)_{34}$ &
$\frac{1}{8}\left[1+\left(\alpha^{*}\beta+\beta^{*}\alpha\right)^{2}\right]$\\
 $|\psi^{-}\rangle_{12}$ &
$\frac{1}{\sqrt{2}}\left(-\beta|00\rangle+\alpha|01\rangle+\alpha|10\rangle-\beta|11\rangle\right)_{34}$ &
$\frac{1}{8}\left[1-\left(\alpha^{*}\beta+\beta^{*}\alpha\right)^{2}\right]$\\
 $|\phi^{+}\rangle_{12}$ &
$\frac{1}{\sqrt{2}}\left(\alpha|00\rangle+\beta|01\rangle+\beta|10\rangle+\alpha|11\rangle\right)_{34}$ &
$\frac{1}{8}\left[1+\left(\alpha^{*}\beta+\beta^{*}\alpha\right)^{2}\right]$\\
 $|\phi^{-}\rangle_{12}$ &
$\frac{1}{\sqrt{2}}\left(\alpha|00\rangle-\beta|01\rangle-\beta|10\rangle+\alpha|11\rangle\right)_{34}$ &
$\frac{1}{8}\left[1-\left(\alpha^{*}\beta+\beta^{*}\alpha\right)^{2}\right]$\\ \noalign{\smallskip}\hline
\end{tabular}
\end{table}

Further we wish to add that the quantum channel chosen by Yang \emph{et al.} are neither robust nor special. For
example, Zhang \emph{et al.} have recently reported a scheme for controlled teleportation by using another
tripartite state
($|\phi\rangle_{Zhang}=\frac{1}{2}\left(|000\rangle+|110\rangle+|101\rangle+|011\rangle\right)$) as quantum
channel. A systematic study \cite{anin-arxic-tele} has revealed that the quantum channels described by Yang
\emph{et al.} and Zhang \emph{et al.} are member of a family of states denoted by
$\frac{1}{\sqrt{2}}\left(\psi^{+}\left|1\right\rangle +\phi^{+}\left|0\right\rangle \right)$ and
$\frac{1}{\sqrt{2}}\left(\psi^{+}\left|0\right\rangle +\phi^{+}\left|1\right\rangle \right)$ respectively.
Altogether there are 12 such channels and essentially the Bell measurement done by Alice swaps the entanglement
and its natural that the Yang's consideration of separable state after the Bell measurement lead to incorrect
conclusions. Further GHZ-\emph{like} states are capable to work as quantum channel for controlled teleportation
of \emph{n}-qubit non-maximally entangled quantum state of the form $\psi=\alpha\left|x\right\rangle
\pm\beta\left|\bar{x}\right\rangle $ where $|\alpha|^{2}+|\beta|^{2}=1$, $x$ varies from $0$ to $2^{n}-1$ and
$\bar{x}=1^{\otimes n}\oplus x$ in modulo 2 arithmetic \cite{anin-arxic-tele}. Thus Yang's scheme only considers
a special case of more generalized scheme proposed in \cite{anin-arxic-tele}.

 %%%%%%%%%%%%%%%%%%%%%%%%%%%%%%%%%%%%%


\begin{thebibliography}{3}
\bibitem{yang-c}K. Yang \emph{et al}., Int. J. Theo.
Phys.\textbf{ 48}, 516 (2009)

\bibitem{zhang-c}Q. Y. Zhang \emph{et al}., Int J Theo.
Phys. \textbf{48}, 3331 (2009)

\bibitem{anin-arxic-tele}A. Banerjee and A. Pathak,
arXiv, quant-ph\textbackslash{}1006.1042 (2010)
\end{thebibliography}
\end{document}